\documentclass{KapProc} 
\usepackage{graphicx}
\let\footnote\savefootnote
\let\footnotetext\savefootnotetext 
\let\footnoterule\savefootnoterule 

\kluwerbib

\startauthorindex

\begin{document}


\articletitle[RX Puppis]
{Circumstellar environment\\ of RX Puppis\thanks{to appear in Post-AGB objects as a 
phase of stellar evolution, eds. R. Szczerba et al., Kluwer}}

\chaptitlerunninghead{RX Puppis}

\author{Joanna Miko{\l}ajewska}
\affil{N. Copernicus Astronomical Center, Bartycka 18, 00716 Warsaw, Poland}
\email{mikolaj@camk.edu.pl}

\author{E. Brandi, L. G. Garc\'{\i}a, O. E. Ferrer} \affil{Facultad de Ciencias 
Astron{\'o}micas y Geof\'{\i}sicas, UNLP - CIC - CONICET,  Argentina} 

\author{P.A. Whitelock, F. Marang} \affil{ South African 
Astronomical Observatory, PO Box 9, 7935, Observatory, South Africa}

\begin{keywords}
Symbiotic binaries, Mira variables, circumstellar matter, mass loss, dust, polarization
\end{keywords}

\begin{abstract} The symbiotic Mira, RX Pup, shows long-term variations in
its mean light level due to variable obscuration by circumstellar dust. The
last increase in extinction towards the Mira, between 1995 and 2000, has
been accompanied by large changes in the degree of polarization in the
optical and red spectral range. The lack of any obvious associated changes in
the position angle may indicate the polarization variations are driven by
changes in the properties of the dust grains (e.g. variable quantity of
dust and variable particle size distribution, due to dust grain formation
and growth) rather than changes in the viewing geometry of the scattering
region(s), e.g. due to the binary rotation.
\end{abstract}


\section{Introduction}

RX Pup is a symbiotic binary composed of a long-period Mira variable
pulsating with $P \approx 578$ days, surrounded by a thick dust shell, and a
hot $\sim 0.8\, \rm M_{\odot}$, white dwarf companion which has been
undergoing a nova-like eruption during the last three decades. The binary
separation could be as large as $a \geq 50$ a.u., and the corresponding
$P_{\rm orb} \geq 200$ yr, as suggested by the permanent presence of a dust
shell around the Mira component (Miko{\l}ajewska et al. 1999, hereafter
M99).  In particular, the Mira is never stripped of its dust envelope, and
even during relatively unobscured phases the star resembles high-mass loss
galactic Miras with thick dust shells.

Our recent analysis of multi-frequency observations shows that most, if not
all, photometric and spectroscopic activity of RX~Pup in the UV, optical and
radio range is due to activity of the hot component, while the Mira variable
and its circumstellar environment is responsible for practically all changes
in the IR range (M99). In particular, we have found large changes in the
reddening towards the Mira accompanied by fading of the near IR flux (Fig.
1).  However, the reddening towards the hot component and emission line
regions remained practically constant and is generally less than that
towards the Mira. These changes do not seem to be related to the orbital
configuration nor to the hot component activity. Similar dust obscuration
events occur in many well observed symbiotic Miras (e.g.  Whitelock 1998),
as well as in some single Miras (e.g. Mattei 1997, Whitelock 1998), and they
are best explained as intrinsic changes in the circumstellar environment of
the Mira variable, possibly due to intensive and variable mass loss (see
also discussion in M99, and Miko{\l}ajewska 1999).

In the following we discuss polarization measurements for RX Pup 
obtained during the most recent obscuration phase.

\section{Observations}

$UBVRI$ polarization measurements were obtained between May 1996 and March
2000 with the five channel photopolarimeter of Turin Observatory, attached
at the 2.15-m CASLEO telescope at San Juan, Argentina.  More details
about the instrumentation and data reduction can be found in Brandi et al. 
(2000).  In addition, optical spectra with the same telescope and REOSC
echelle spectrograph were obtained (see M99 for details).  Finally, new
$JHKL$ broad-band photometry on the SAAO system was obtained with the Mk\,II
infrared photometer on the 0.75-m telescope at Sutherland (Carter 1990). 
These new IR data together with data from M99 are displayed in Figure 1
where the dates of the polarimetric measurements are also marked.

\section{Results and discussion}

In the period covered by our polarimetric observations, the near IR flux of
RX Pup was gradually decreasing and the $J-K$ colour was increasing,
suggesting that the Mira had entered a new obscuration phase. These secular
changes in near IR flux are generally not correlated with changes in the
optical; in particular the visual magnitude estimates reported by the
Variable Star Section of the Royal Society of new Zealand (RASNZ) show that
RX Pup was brightening in the optical between 1995 and 1998, and declining
later (see also Fig.\ 4 in M99).

\begin{figure}[ht]
\begin{minipage}[b]{0.65\textwidth}
\includegraphics[width=\textwidth]{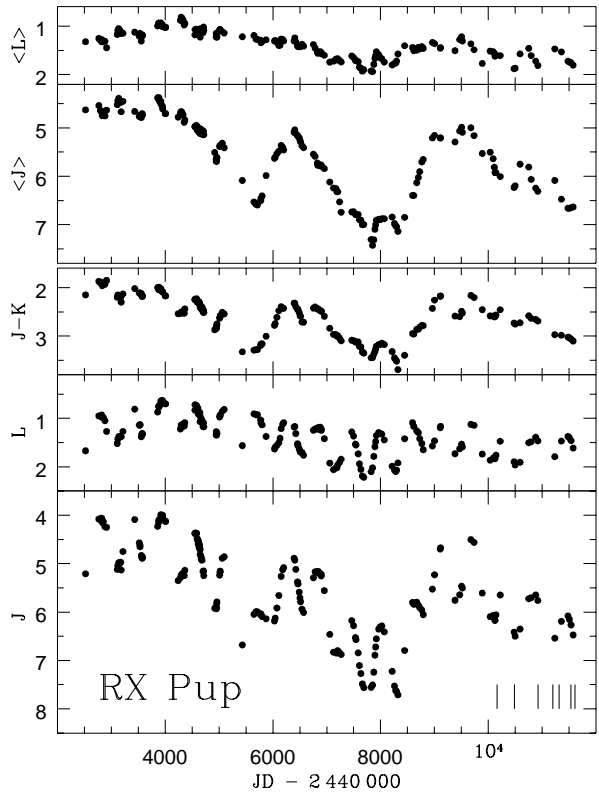}
\end{minipage}
\hfill
\begin{minipage}[b]{0.3\textwidth}
\caption{$J$ and $L$ light curves and $J-K$ colour of RX Pup. The upper panel
show the same light curves after removal of the Mira pulsation.  Bars in the
bottom panel indicate dates of our polarimetric observations.}
\end{minipage}
\end{figure}

The results of polarimetric observations are shown in Figure 2.  The left
panel presents the degree of polarization and the position angle as a
function of wavelength for different observing runs.  The observed
wavelength dependence of $p$ and the position angle could be accounted for
by the presence of two components, one with p.a. $\sim 145-155^{\circ}$, and
a flat wavelength dependence resembling that of interstellar polarization
which suggests a presence of interstellar-type large dust particles, and
another component with p.a. $\sim 175-160^{\circ}$ and $p$ increasing
steeply towards the ultraviolet, consistent with Rayleigh scattering by
small particles. Both components show significant temporal changes which
indicate circumstellar origin of the observed polarization.

Figure 2 also shows the polarization changes versus the visual magnitudes
and $J-K$ colour. The visual magnitude is dominated by the light from the
hot component, whereas the $J-K$ colour measures the amount of reddening
towards the Mira component. The polarization degree seems to increase with
both the visual magnitude and the $J-K$ colour. At the same time we do not
find any correlation between the polarization and the Mira pulsation phase.

The polarized flux does not seem to be scattered light from the Mira, since
in the optical the Mira is heavily obscured by the dust (M99). In
particular, we do not see pulsational variations, nor can we
detect the TiO bands in optical spectra at any epoch (see also M99). The
$UVBR$ and probably also the $I$ flux is dominated by the hot component
with some contribution from the nebular line and continuum emission.  M99
also found that the contribution from the nebular emission increases
relative to other sources of emission as the optical brightness decreases,
and as the hot component temperature and degree of ionization in the nebula,
increase. In particular, the contribution from the nebular emission was
negligible in 1995-1996. Then the relative contribution of the nebular
emission increased from mid 1998 in-step with the declining visual
brightness, although it never exceeded 0.1-0.15 mag. The polarized flux
varied by a factor of $\sim 2$ in the visual range, whereas the strongest
H\,{\sc i} Balmer and Fe\,{\sc ii} emission-line fluxes remained constant
within $\sim 20\%$.  There was also no significant increase in polarization
(although only $p_{\rm R}$ estimates are available) in 1987, when the
nebular continuum and emission lines where much stronger than in the 1990s
(M99). It is thus unlikely, that the polarized component corresponds to the
nebular emission. Instead we believe that it is most likely to be radiation
from the warm, $T \sim 6000\, \rm K$, component found in the 1995-96 optical
spectra by M99 -- scattered in the dust envelope surrounding the Mira.
The warm component is responsible for the optical variability in
the 1990s; however, there is no obvious, simple correlation between this
activity and the reported polarization changes.

\begin{figure}[t]
\includegraphics[width=2.3in]{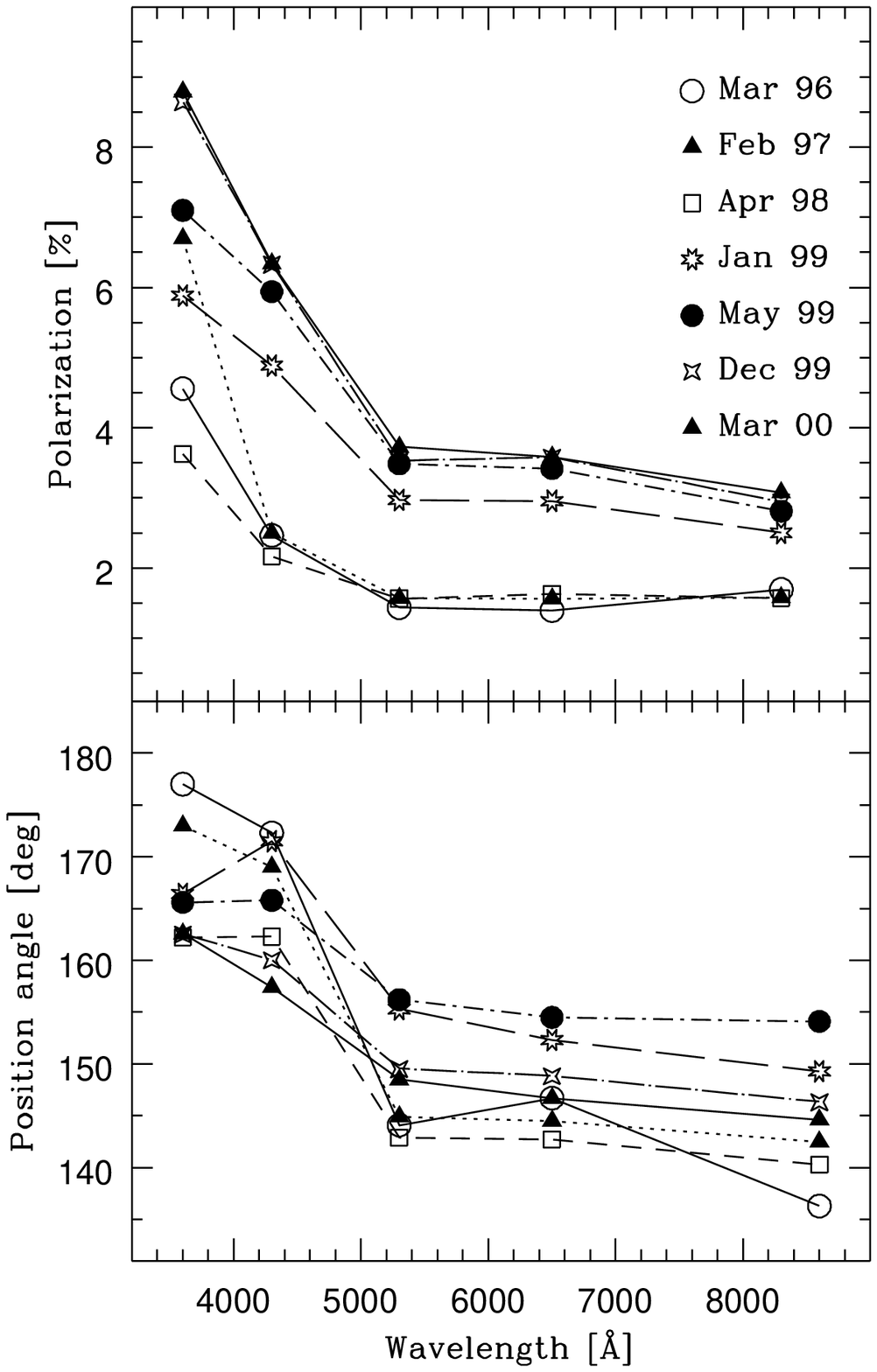}
\hspace{0.2in} 
\includegraphics[width=2.2in]{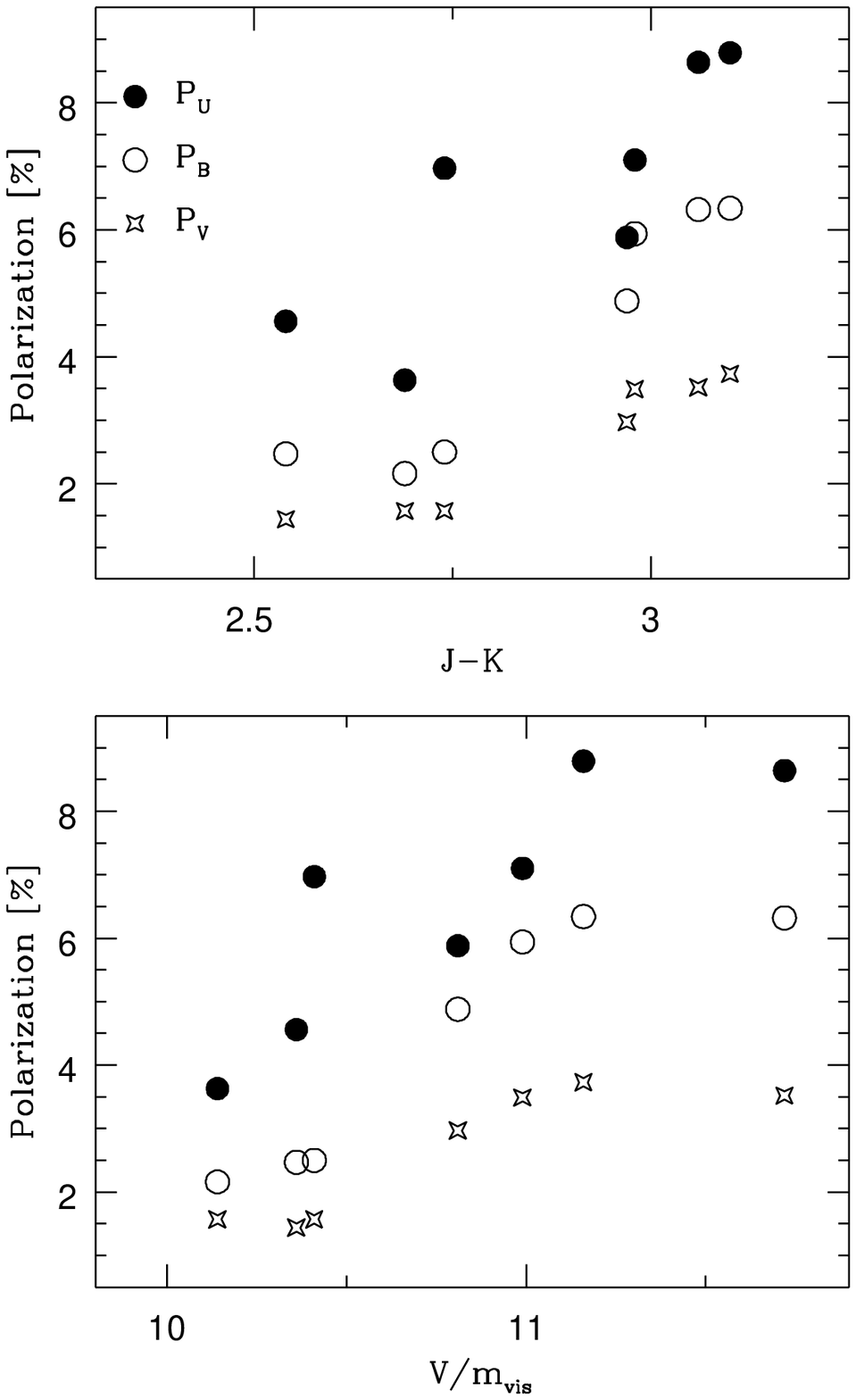} 
\caption{Left: Wavelength dependence of linear polarization, $p\,[\%]$ (top), and 
position angle for different epochs (bottom). The points correspond to weighted mean 
values calculated for each observing run. The error bars are not shown, however the 
relative mean errors in $p\,[\%]$ are $< 0.06 $ in $UB$,  $< 0.008$ in $VRI$, and $< 
0.01$ in the position angle. 
Right: Dependence of the
polarization on the reddening towards the Mira measured by the $J-K$ colour
(top), and visual brightness (bottom).}
\end{figure}

The changes in polarization are apparently correlated with the varying
reddening towards the Mira. Between 1994/5 and 1999/2000 the Mira's average
brightness weakened by $\sim 1.5$ mag at $J$ and $\sim 0.7$ mag at $K$,
respectively. The mean $J-K$ colour increased from $\sim 2.2$ in 1994/5 to
$\sim 2.6$ in 1996--98, and to $\sim 3$ in 1999, respectively. These changes
have been accompanied by an increase in $p$, by a factor of $\sim 2-3$, at
all wavelengths, and a slow rotation in the position angle. The sense of
this rotation, however, depends on the spectral range: the position angle
decreased by $\sim 15^{\circ}$ in $UB$ and increased by $\sim 10^{\circ}$ in
the $VRI$ range, respectively. Unfortunately, it is practically impossible to
distinguish between a steady rotation of position angle with time and
changes related to variable conditions in the circumstellar envelope of RX
Pup because the reddening has been steadily increasing during the
period covered by our observations.

There are very few previous polarization measurements for RX Pup. The $UBV$
polarimetric measurements obtained during maximum of the hot component
outburst and IR bright phase of the Mira in 1979-1980 show practically flat
wavelength dependence with $p \sim 2 \%$ and a constant position angle of
$\sim 120^{\circ}$ (Barbier \& Swings 1982); however, these data 
were corrected, by an unknown amount, for interstellar polarization.  In any
case, the component with $p$ steeply increasing towards ultraviolet was
absent at this epoch. Schulte-Ladbeck et al. (1990) estimated $p_{\rm R} =
1.61$ and ${\rm p.a.} = 121^{\circ}$ on Mar 27, 1987, whereas Harries \& Howarth
(1996) measured $p_{\rm R} = 1.825$ and ${\rm p.a.} = 146.7^{\circ}$ on May
6-10, 1994. On both of these dates, the visual magnitude of RX Pup was
$V/m_{\rm vis} \sim 11.1$ and $J-K \sim 2.5$.  We have observed similar
values of $p_{\rm R} \sim 1.5-1.6$ and $J-K \sim 2.5-2.7$ in 1996-98; the
system was then, however,  much brighter, $V \sim 10.4-10.1$. This may
suggest that the polarization changes are related predominantly to changes
in the reddening towards the Mira, and not to the hot component brightness.
The polarization changes are also not related to any particular changes in
the emission line spectrum of RX Pup.

Interestingly, during 1996 March to 1998 April the degree of polarization at
$BVRI$ was practically constant, while in $U$ it varied. This may indicate
that the 1996-98 changes were caused by variable conditions in the region
where small particles dominate, while in 1999-2000 the contribution of large dust
particles becomes comparably important. The lack of dramatic changes in the
position angle associated with changes in $p$ and the reddening towards the
Mira component may indicate the polarization variations are driven by
changes in the properties of the dust grains, for example variable amounts
of dust with variable size distributions, due to dust grain formation and
growth.

If the rotation in position angle with time is due to orbital motion, than
the change in p.a. of $\sim 10 - 15^{\circ}$ in $\sim 5$ years corresponds
to the binary period of $\sim 120 - 180$ years. Similarly, $\Delta\,\rm p.a.
\sim 30$ between 1987 and the most recent observations is consistent with $P
\sim 160$ yr. Note that an orbital period of $\sim 200$ yr is also implied
by the properties of the dust envelope around the Mira component (M99).

In this context, it is interesting that a similar increase in $p$ at
$UBVRI$, together with  a slow rotation of the position angle,
increasing in $UB$ and decreasing in $RI$, was observed during the dust
obscuration event in R Aqr in the late 1970s (Nikitin \& Khydyakova 1979).
There is, however, a significant difference between R Aqr and RX Pup; in the
former the optical and red light is dominated by the Mira, whereas in the
latter the contribution from the Mira is negligible at these wavelengths .
Moreover, in R Aqr, both the degree of polarization and the position angle
(especially in the ultraviolet and blue spectral range) also vary with
pulsation phase, which is not the case for RX Pup.

Since we do not know the orbital phase of RX Pup, it is impossible to drive
any definite conclusions from the position angle of the polarized light.
However, the fact that the hot component does not seem much affected when
the reddening towards the Mira is strongly increasing (M99), suggests that
it cannot be behind the dust cloud. Furthermore, the relatively large degree
of polarization suggests that it is to one side of, rather than in front of,
the dust.  Assuming that the dust around the Mira is largely confined to a
cone shaped region, the hot star is not far from the quadrature and is
moving in our direction (somewhere between the quadrature and the inferior
conjunction). According to the standard scattering model the position angle
of the polarized light should be roughly orthogonal to the line connecting
the source of light and the scattering region. The observed position angle
short wavelengths, where the Rayleigh scattering by small particles
dominates, is then roughly consistent with a scattering region close to the
Mira, while the large particles are presumably dominating at larger
distances from the Mira, which may account for the difference between
p.a. at $UB$ and $VRI$.

The polarization geometry of RX Pup can be compared with the geometry of its
ionized nebula. There are two main directions for mass outflow distinguished
in the nebula which are roughly perpendicular to each other (Corradi \&
Schwarz 2000): the previously known elongated optical/radio feature at p.a.
$\sim 15^{\circ}$ with a velocity decreasing with distance from the central
object, and a new EW compact component expanding at $\geq 80$ km/s, probably
with a bipolar shape. The elongated structure would then be roughly
perpendicular to the deduced binary axis, whereas the compact bipolar flow
would be aligned with the binary axis.

The orientation of the binary axis will of course change with the binary
motion. The shape (geometry) of the neutral region where the dust can be
effectively formed, and so the scattering geometry, will also change with
the hot component properties, for example with temperature and luminosity -
which determine the number of ionizing photons available, and with wind
efficiency - which influences the position of the shock front between the
components. Thus, further studies involving both high resolution imaging
techniques together with spectropolarimetry are necessary to reveal much
more detailed information about the geometry of the circumstellar
environment of RX Pup.


\begin{acknowledgments}
This research was partly supported by KBN Research Grants No. 2\,P03D\,021\,12 and No. 
5\,P03D\,019\,20 and makes use of observations obtained at the SAAO,
Sutherland, South Africa.
\end{acknowledgments}


\end{document}